\documentclass[aps,twocolumn,superscriptaddress,preprintnumbers,article]{revtex4}

\usepackage[latin9]{inputenc}
\usepackage{color}
\usepackage{verbatim}
\usepackage{amsmath}
\usepackage{amssymb}
\usepackage{graphicx}
\usepackage{esint}
\usepackage{esvect}
\usepackage{braket}
\usepackage{graphicx}
\usepackage{subfigure}
\usepackage{hyperref}

\makeatletter
\@ifundefined{textcolor}{}
{%
 \definecolor{BLACK}{gray}{0}
 \definecolor{WHITE}{gray}{1}
 \definecolor{RED}{rgb}{1,0,0}
 \definecolor{GREEN}{rgb}{0,1,0}
 \definecolor{BLUE}{rgb}{0,0,1}
 \definecolor{CYAN}{cmyk}{1,0,0,0}
 \definecolor{MAGENTA}{cmyk}{0,1,0,0}
 \definecolor{YELLOW}{cmyk}{0,0,1,0}
}

\usepackage{mathrsfs}

\draft
\makeatother
\begin{document}

\title{A Scaling Behavior of Bloch Oscillation in Weyl Semimetals}
\author{Yan-Qi Wang}
\affiliation{International Center for Quantum Materials,School of Physics, Peking University, Beijing 100871, China}
\affiliation{Collaborative Innovation Center of Quantum Matter, Beijing 100871, China}
\author{Xiong-Jun Liu \footnote{Corresponding author: xiongjunliu@pku.edu.cn}}
\affiliation{International Center for Quantum Materials,School of Physics, Peking University, Beijing 100871, China}
\affiliation{Collaborative Innovation Center of Quantum Matter, Beijing 100871, China}

\begin{abstract}
We predict a linear logarithmical scaling law of Bloch oscillation dynamics in Weyl semimetals (WSMs), which can be applied to detect Weyl nodal points. Applying the semiclassical dynamics for quasiparticles which are accelerated bypassing a Weyl point, we show that transverse drift exhibits asymptotically a linear log-log relation with respect to the minimal momentum measured from the Weyl point. This linear scaling behavior is a consequence of the monopole structure nearby the Weyl points, thus providing a direct measurement of the topological nodal points, with the chirality and anisotropy being precisely determined. We apply the present results to two lattice models for WSMs which can be realized with cold atoms in experiment, and propose realistic schemes for the experimental detection. With the analytic and numerical results we show the feasibility of identifying topological Weyl nodal points based on the present prediction.
\end{abstract}
\pacs{71.10.Pm, 73.50.-h, 73.63.-b}
\date{\today }
\maketitle

{\it Introduction.}--Weyl semimetals (WSMs), being a new type of free fermion topological states~\cite{Wan2011,Xu2011,Burkov2011,Hosur2012,Jiang2012}, have attracted fast growing attention since the experimental discovery in the solid state material TaAs in last year~\cite{HASAN2,DAIXI1,DAIXI2}. Differently from topological insulators which have gap in the bulk and gapless edge or surface modes in the boundary~\cite{HASAN,QI}, WSMs exhibit gapless bulk nodal points, called Weyl points, in the Brillouin zone~\cite{Nielsen1981,Nielsen1983}. Due to the absence of a bulk gap, the topology of a WSM is defined not on the three-dimensional (3D) bulk bands, but on the 2D gapless Fermi surfaces enclosing Weyl points in the Brillouin zone~\cite{Wan2011,Xu2011,Burkov2011,Hosur2012,Jiang2012}. It follows that such nodal points are topologically protected, and have chiralities which are characterized by Chern numbers and figured as monopoles in momentum space. A number of fascinating physics have been predicted for WSMs, including surface Fermi arc states~\cite{Wan2011,Xu2011,Burkov2011} and different magneto-transport anomalies~\cite{Nielsen1983,Hosur2012,Zyuzin2012,Son2013,Parameswaran2014,CZhang2015,Xiong2015,Jiang2015,FZhang2015}, with some of them having been observed in the recent experiments~\cite{HASAN2,DAIXI1,DAIXI2}. Nevertheless, to observe directly the bulk topology of WSMs, including the chirality and monopole structure of Weyl points, are usually not realistic in solid state materials.

Recently, fueled by the great progress in the experimental realization, the synthetic spin-orbit (SO) coupling~\cite{Liu2009,Lin,Jinyi2012,MIT,Wang} and gauge fields~\cite{Abeliangauge2,Abeliangauge3,Abeliangauge4} have developed into a most active area for cold atoms~\cite{GJOS,HZhai2015}. In the presence of synthetic SO coupling or gauge fields, many novel topological states have been proposed and studied for cold atoms~\cite{WULIU,Goldman2012,Liu2013a,Vincent2013,Chuanwei2013,LIU2,Luming,HChen2015,Hickey2016}, including the 3D Weyl semimetals in optical lattices~\cite{KETTERLE,HE}. Compared with solid state materials, the cold atoms may offer advantages to the detection of topological states, e.g. through the direct measurement of the bulk states which can hardly be observed for solids~\cite{Alba2011,Price2012,LP,Dongling2014,LIU3}. Note that by definition a possible standard way to observe bulk topology of a WSM is to measure the integral of Berry curvature throughout a 2D closed surface in Brillouin zone enclosing the Weyl point. This, however, is generically a very challenging task for real cold atom experiments, making the direct detection of topological Weyl points be still an unresolved important question.

In this work, we predict a linear logarithmical scaling law of Bloch oscillation dynamics in WSMs and propose to detect the Weyl nodal points based on this universal behavior. From the semiclassical dynamics of quasiparticles which are accelerated close to a Weyl point, we show that transverse drift exhibits a linear log-log relation with respect to the minimal momentum measured from the Weyl point. The monopole Berry curvature, chirality, and anisotropy of the Weyl points can be read out from the predicted scaling behavior. We apply the present results to lattice models for WSMs which can be realized in cold atoms, and show the feasibility of experimental measurements by analytic and numerical studies.

\emph{Continuous model.}--We start with the continuous model for a generic WSMs, which can be type I or type II~\cite{BD,HCV,EXTYPE21}. The continuous Hamiltonian nearby a Weyl point reads (taking $\hbar=1$)
\begin{equation}\label{drift1}
{\cal H}(\bold k)=v_0k_x  \otimes {\bf 1} +\sum_{j=x,y,z}v_jk_j\sigma_j,
\end{equation}
where ${\bf 1}$ is a $2 \times 2$ unit matrix, the coefficients $v_0$ and $v_j$ determine the velocity of quasiparticles along the $j=x,y,z$ direction, and $\sigma_{x,y,z}$ are Pauli matrices. It is trivial to see that for $v_0\neq0$, the velocity of quasiparticles is symmetric with respect to $y$ and $z$ directions, but asymmetric in the $x$ direction. In particular, when $|v_0|>|v_x|$, the $x$ component of the velocity is always positive (for $v_0>0$) or negative (for $v_0<0$), rendering a type II WSM~\cite{BD,HCV,EXTYPE21}. Denote by $\ket{u_{-, {\bf k}}}$ the eigenstates for the lower subband. The Berry connection $\mathcal{A}_{-}{\bf(k)} = i\bra{u_{-,{\bf k}}} \nabla_{{\bf k}} \ket{u_{-, {\bf k}}}$ and Berry curvature $ \mathcal{\vec B}_{-}({\bf k}) = \nabla_{{\bf k}} \times \mathcal{A}_{-}{\bf (k)}$, giving that ${\cal B}_-^i ={v_x v_y v_zk_i}/{[2(\sum_jv_j^2 k_j^2)^{3/2}]} \quad {(i = x, y ,z)}$.
The topology of the Weyl point is characterized by the first Chern number, which is calculated by the integral of Berry curvature throughout a surface enclosing the Weyl point at $\bold k=0$, namely $C_{\bf 1}=\frac{1}{2\pi} \oint d{\bf S} \cdot{} \mathcal{\vec B}_{-}({\bf k})$. Direct calculation shows that $C_{\bf 1} = 1$, reflecting the chirality of the Weyl point, and a monopole located at $\bold k=0$.

We study Bloch oscillation of a Bose-Einstein condensate (BEC) prepared at lower subband with center-of-mass position ${\bf r}$ and momentum ${\bf k}$ by applying an external force $\vec F$~\cite{LP}. The semiclassical dynamics is governed by $\dot{k}_j = F_j$, $\dot{r}_j =  \partial_{{k}_j} {\cal E}_-({\bf k}) + \epsilon_{jkl}F_k{\mathcal{B}}_{-,l}$, where ${\cal E}_-({\bf k})=v_0k_x-(\sum_jv_j^2k_j^2)^{1/2}$. The transverse drift induced by Berry curvature of monopole is calculated by $S = -\int \vec F \times \mathcal{\vec B}_-({\bf k}) dt = -\int d {\bf k} \times \mathcal{\vec B}_-({\bf k})$. This transverse drift is independent of strength of force but sensitive to magnitude of Berry curvature. It is easy to see that the transverse drifts induced by two Weyl points with opposite chiralities are in the opposite directions. Such a direction provides a simple measurement of the chirality of Weyl points. Without loss of generality, we consider that the initial momentum of the BEC is $\bold k_0= k_0 \hat e_n =k_0\cos \theta_0 \cos \phi \hat e_x+ k_0\cos \theta_0 \sin \phi\hat e_y -k_0\sin \theta_0\hat e_z $, where $\hat e_n$ is a unit vector with $\theta_0 + \pi/2$ and $\phi$ being its polar and azimuthal angles in the spherical coordinate constructed on the Weyl point. A generic force can be described by $\vec F = -F\cos(\theta_0 + \theta) \cos \phi\hat e_x -F\cos(\theta_0 + \theta) \sin \phi\hat e_y+F \sin (\theta_0 + \theta)\hat e_z$. Here $\theta$ is the angle between $\vec F$ and line from $\bold k_0$ to the Weyl point. The transverse drift, after a long enough time evolution, is integrated out via $S(t\rightarrow\infty)=-\int_0^\infty \vec F \times \vec{\mathcal{B}}({\bf k}) dt$ that
\begin{eqnarray}\label{drift2}
S{[t\rightarrow\infty]} =\frac{v_x v_y(\sqrt{\sum_i {(\hat e_n \cdot{} \hat e_i)}^2v_i^2}   + P(\theta))}{v_zk_0 \sin\theta(v_x^2 \cos^2 \phi + v_y^2 \sin^2 \phi)}
\end{eqnarray}
where $P(\theta) = \left\{ (v_x^2\cos^2\phi \right.$$ + v_y^2 \sin^2\phi)$$[(\cos \theta_0 + \cos (\theta_0 + \theta))$$\cos (\theta_0 + \theta)$$- 2\cos(\theta_0)^2] + v_z^2[(\sin(\theta_0 + \theta)$$ + \sin \theta_0)\sin(\theta_0 + \theta)$$\left.- 2\sin^2(\theta_0)] \right\} [4\sum_i {(\hat e_n \cdot{} \hat e_i)}^2v_i^2 ]^{-\frac{1}{2}}$ is a function of $\theta$. The above formula approaches a simple relation in the small angle limit $\theta \rightarrow0$ that: $P(\theta) = 0$ and
$k_{\text{min}}S={v_x v_y\sqrt{\sum_i {(\hat e_n \cdot{} \hat e_i)}^2v_i^2} }/{v_z(v_x^2 \cos^2 \phi + v_y^2 \sin^2 \phi)}$, with $k_{\text{min}} = k_0 \sin \theta$ the minimal momentum measured from the Weyl point in the whole dynamical process. With this result, we reach a linear log-log scaling by
\begin{eqnarray}\label{drift3}
\ln S = -\ln k_{\text{min}} + \ln \frac{v_x v_y\sqrt{\sum_i {(\hat e_n \cdot{} \hat e_i)}^2v_i^2}}{v_z(v_x^2 \cos^2 \phi + v_y^2 \sin^2 \phi)}, \  \theta\rightarrow0.
\end{eqnarray}
The linear scaling relation is deeply rooted in the fact that in the small $\theta$ limit, $S$ is dominated by the monopole Berry curvature close to the Weyl point, which exhibits the inverse square law $|{\cal B}_-|\propto1/k^2$ with respect to the momentum. Thus this log-log scaling behavior is universal and model independent, as we shall further prove it in the lattice models.
In particular, if we set $\theta_0 = 0$ and $\phi = \pi/2$, (i.e. $\bold k_0 = -k_0\hat e_y$), the intercept of the above formula~\eqref{drift3} gives $D_x = \ln (v_x/v_z)$, which measures the anisotropy of Weyl point with respect to the two ($x$ and $z$) directions perpendicular to the applied force, and vanishes if $v_x=v_z$. In a similar way, we obtain that $D_y = \ln (v_y/v_x)$ if the initial momentum is $\bold k_0 = -k_0\hat e_z$, and $D_z = \ln(v_z/v_y)$ for the initial momentum $\bold k_0 = -k_0\hat e_x$. With these results we can read out the complete information of a topological Weyl point.

\begin{figure}[t]
\centering
\includegraphics[width=0.95\columnwidth]{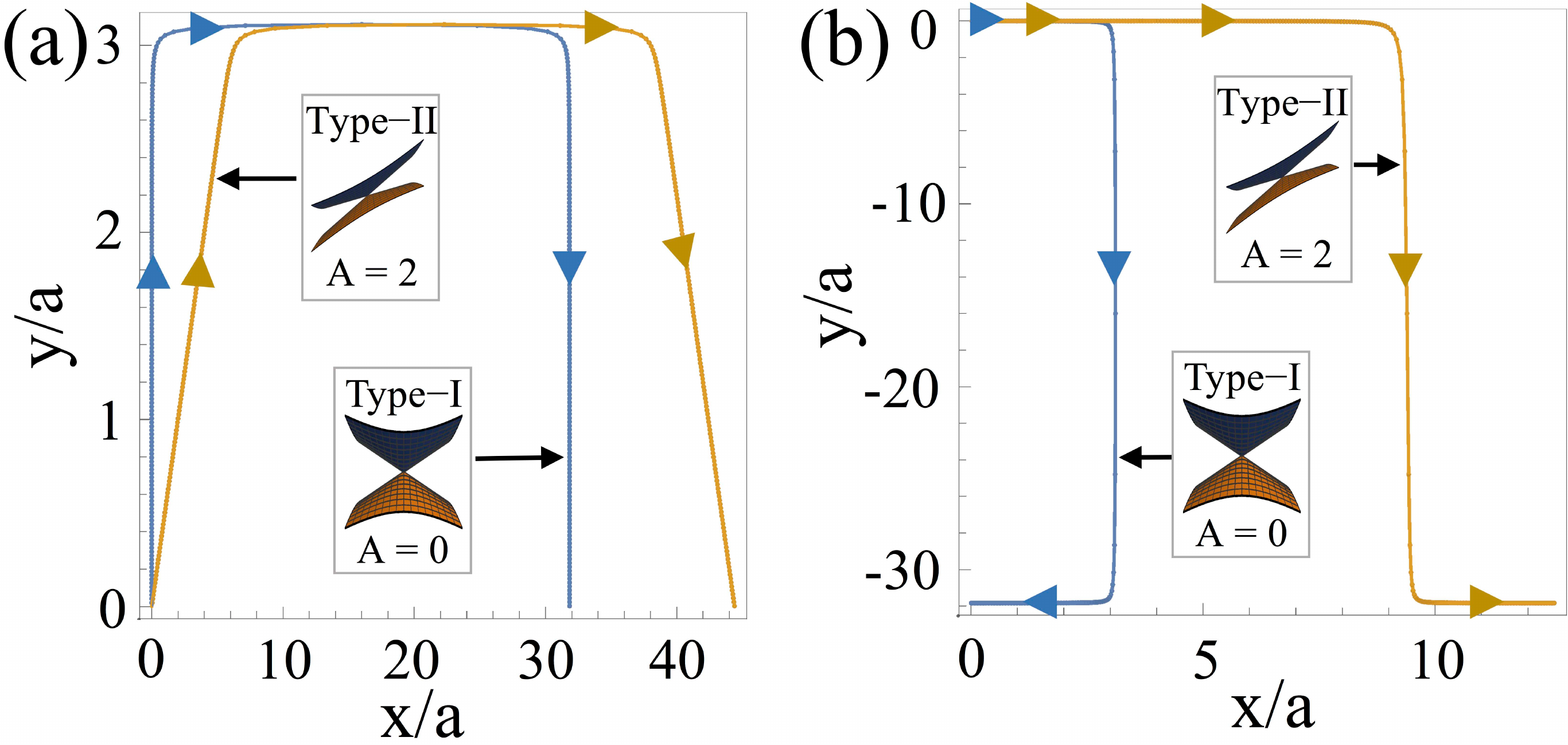}
\caption{\label{typeII} Trajectory of BEC cloud for type I and type II WSMs. (a) Transverse drift occurs
in $x$ direction, with parameters $\bold k_0 = -k_0\hat e_y, F_x = 0, F_y = F \cos\theta$, and $F_z = F \sin\theta$.
(b) Transverse drift occurs in $y$ direction with parameters $\bold k_0 =-k_0\hat e_x, F_x = F \cos\theta, F_y = 0,$ and $F_z = F \sin\theta$.}
\end{figure}
It is interesting that the above scaling behavior is valid for both type I and type II WSMs, while the two different types of WSMs can be distinguished in experiment by the trajectories of BEC cloud accelerated bypassing the Weyl point. As shown in Fig.~\ref{typeII}, the projection of the BEC trajectory onto $x$ axis keeps the same (reverses) direction before and after the particle passes by the Weyl point of the type II (type I) WSM. This is because in such process the sign of the $x$-component normal velocity $\partial{\cal E}_-/\partial k_x= v_0 - {v_x^2k_x}/{\sqrt{\sum_jv_j^2 k_j^2}}$ keeps (changes) sign for the type II (type I) WSM.

\emph{Scaling law in lattice models.}--We proceed to study the scaling behavior in lattice models. Differently from the continuous model, in optical lattices the transverse drift shall be obtained after the BEC cloud completes a Bloch oscillation period. Below we consider two typical lattice models for WSMs which can be realized by generating synthetic gauge fields and SO coupling, respectively.
 \begin{figure}[t]
\centering
\includegraphics[width=0.95\columnwidth]{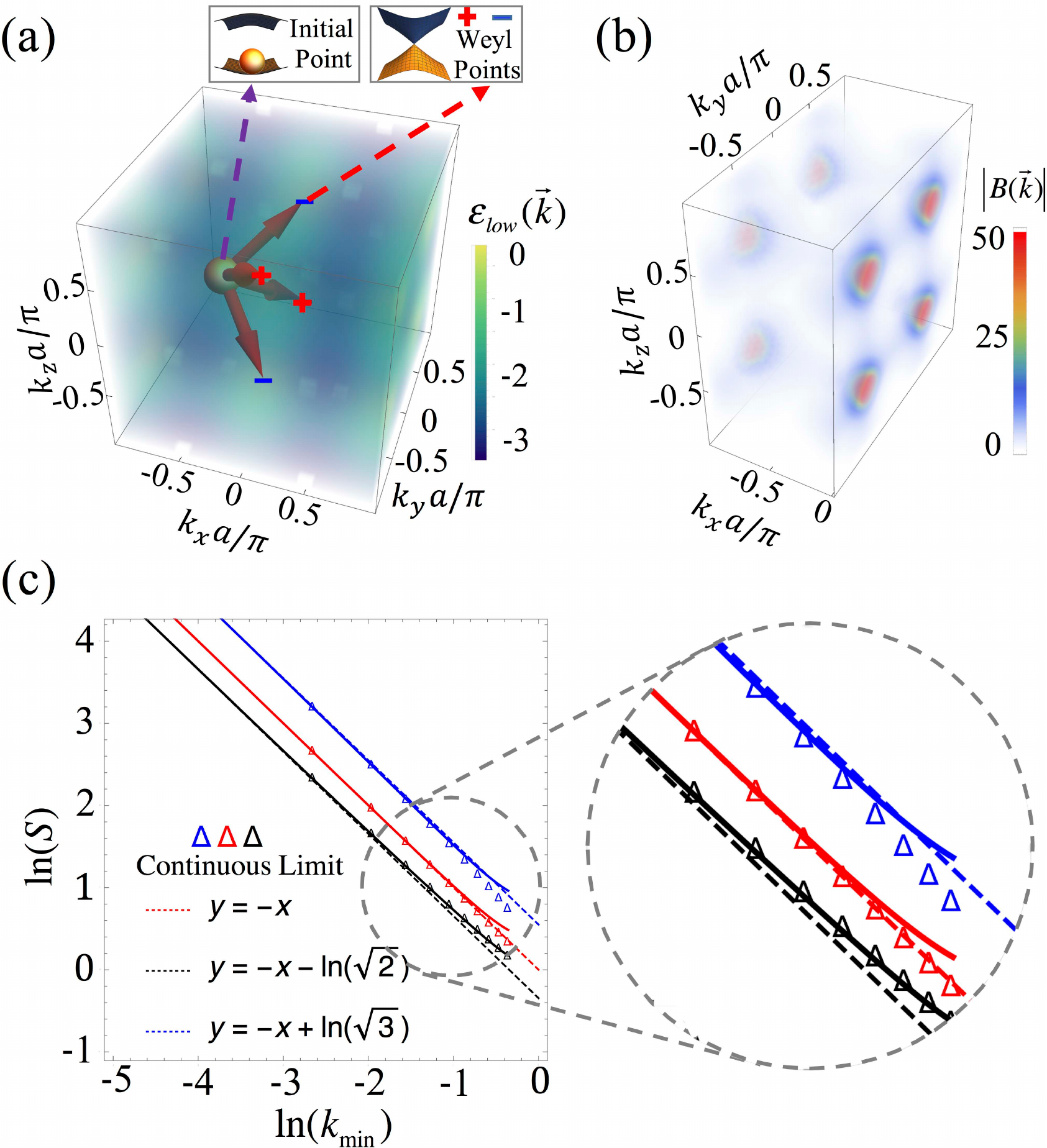}
\caption{\label{figure1} WSM realized with LAT scheme. {(a)} Schematic diagram for BEC cloud accelerated from initial momentum toward Weyl pints; {(b)} Magnitude of Berry curvature. Both (a) and (b) the parameters are taken as with $J_y=K_z=K_x=1.0$, $|\mathcal{\vec B}_{-}({\bf k}) |$ larger than 50 being plotted in red color; {(c)} Scaling relation between drift $S$ and $k_{\rm min}$, with parameters $2J_y=K_z = K_x$ (upper, blue lines), $J_y=K_z=K_x$ (middle, red lines), and $0.5J_y=K_z=K_x$ (lower, black lines). The triangle, solid, and dotted lines represent results in the continuous limit, lattice model, and asymptotic scaling lines. All parameters are rescaled to be dimensionless.}
\end{figure}

The first spinless model for our investigation was proposed by T. Dub\v{c}ek et.al, who realize a Weyl semimetal in a cubic lattice, formed by two ($A$ and $B$) sublattices, by generating synthetic gauge potentials by the laser-assisted tunneling (LAT) scheme~\cite{KETTERLE}. The Bloch Hamiltonian in their realization takes the form
 \begin{eqnarray}\label{LAT}
 \mathcal{H}({\bf k}) = &-2 J_y \cos (k_ya) \tau_x -2K_x \sin (k_xa)\tau_y\nonumber \\
 &+2K_z \cos(k_z a) \tau_z,
 \end{eqnarray}
where the Pauli matrices $\tau_{x,y,z}$ act on the pseudospin (sublattice) space, $(K_x,J_y)$ denote the tunneling amplitudes between $A$ and $B$ sites, and $K_z$ represents the $AA$ and $BB$ hopping along z direction~\cite{KETTERLE}. The lattice constant can be set as $a = 1$ to facilitate further discussions.

Note that the Hamiltonian~\eqref{LAT} preserves time-reversal (TR) symmetry, which is defined as ${\cal T}={\cal K}$ with ${\cal K}$ being the complex conjugate for spinless system. With the TR symmetry, the above system has four independent Weyl points at $\{\bold k_w\}=(0,\pm \pi/2,\pm \pi/2)$, nearby which the Hamiltonian can be linearized as $\mathcal H{\bf (k)} = -2K_x k_x\sigma_y \pm 2J_y k_y\sigma_x  \pm 2K_z k_z\sigma_z$.
This formula corresponds to the case with $v_0=0$ in the equation~\eqref{drift1}. The magnitude of Berry curvature is shown in Fig.~\ref{figure1} (b), reflecting the monopole structures around the Weyl points.

The band bottom is located at four different momenta of the Brillouin zone $\{\bold k_0\}=\{(\pm\pi/2, 0,0),(\pm\pi/2, 0,\pi)\}$. To study the scaling relation, we consider that the BEC cloud is prepared at $\bold k_0=(-\pi/2, 0,0)$ and then accelerated by an external force toward one of the four Weyl points. For this purpose the force can be applied in the $\hat e_\pm-\hat e_y$ plane, with $\hat e_\pm=(\hat e_x\pm\hat e_z)/\sqrt{2}$. Specifically, one has $F_x = (F/\sqrt{2})\cos(\theta_0 + \theta) $, $F_z=\pm (F/\sqrt{2})\cos(\theta_0 + \theta)$, and $F_y =\pm F\sin(\theta_0+\theta)$, where $\theta_0 = \arcsin \sqrt{1/3}$. When $\theta=0$ the force points from the initial momentum to one of the four Weyl points. Under this configuration and in the limit $\theta\rightarrow0$, the linear scaling relation reads
\begin{eqnarray}
\ln S=-\ln k_{\rm min}+\ln \frac{ 2\left| K_xK_z\right| \sqrt{K_x^2+J_y^2+K_z^2}}{\sqrt{3}\left|J_y(K_x^2+K_z^2)\right|}.
\end{eqnarray}
In the experiment, the two LAT couplings $K_x$ and $K_z$ can be tuned as equal, say $K_x = K_z = K$~\cite{KETTERLE}. In this case the intercept can be simplified by
\begin{equation}
D_{\rm LAT}=\frac{1}{2} \ln\bigr[\frac{1}{3} (1+2\frac{K^2}{J_y^2})\bigr].
\end{equation}
Numerical results for the lattice model are shown in Fig.\ref{figure1}~(c). It can be seen that when the lattice is isotropic in all three directions, i.e. $J_y=K$, the intercept vanishes $D_{\rm LAT}=0$ (red curves), which is consistent with the prediction based on the continuous model. In contrast, a nonzero intercept appears for the linear scaling line when the Weyl points are anisotropic. Moreover, from Fig.\ref{figure1}~(c) one finds that the relation between drift $S$ and $k_{\rm min}$ approaches quickly the linear log-log relation when $k_{\rm min}$ is less than $0.1$. The numerical simulation shows that the Weyl points, including their anisotropy, can be well detected in the realistic cold atom experiments.

\begin{figure}[t]
\centering
\includegraphics[width=0.95\columnwidth]{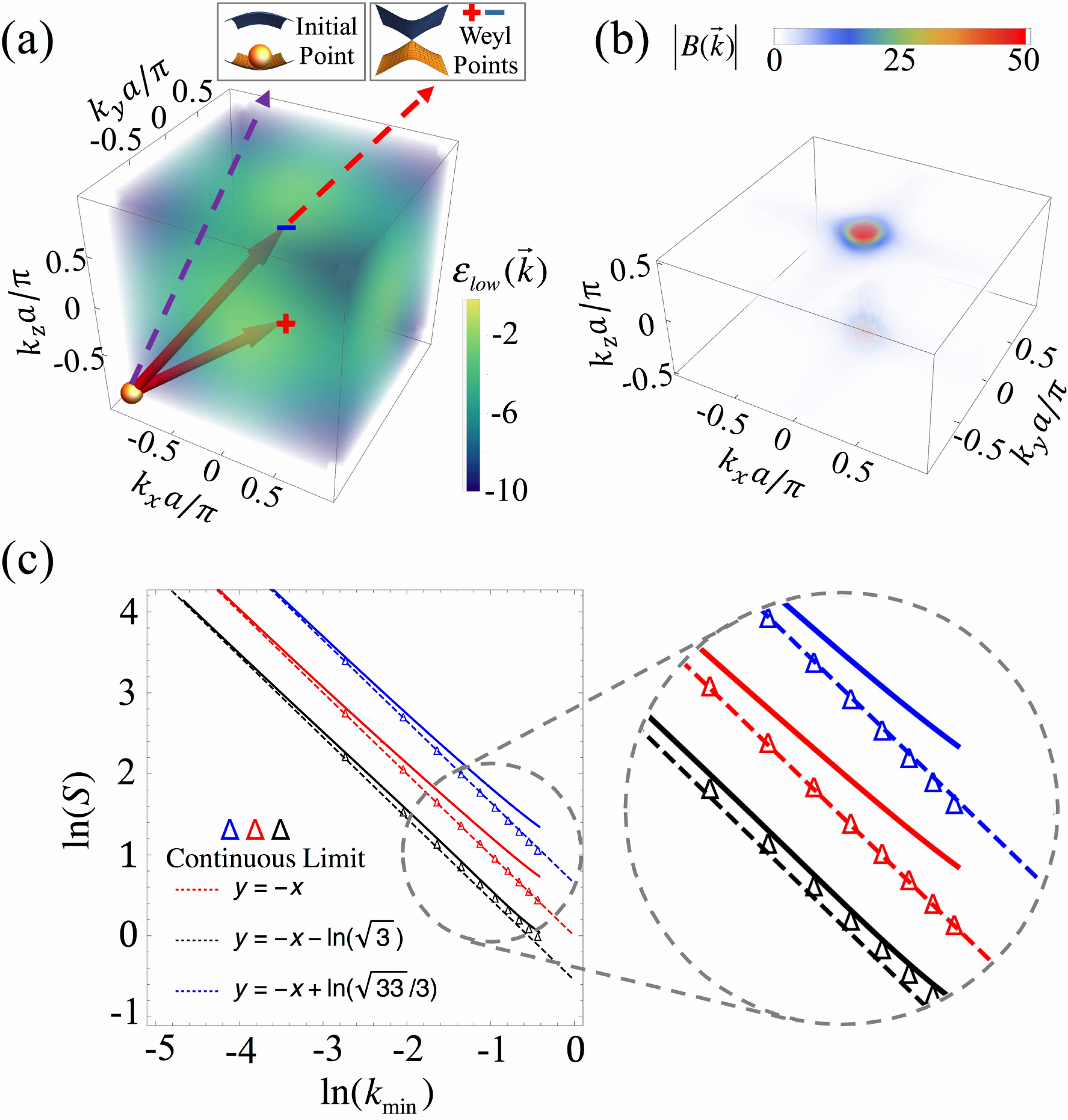}
\caption{\label{figure2} WSM realized with SO coupled lattice. {(a)} Schematic diagram for BEC cloud accelerated from initial momentum toward Weyl pints ; {(b)} Magnitude of Berry curvature. The parameters $t_{\rm so} =t_z= 1.1t_0$, $m_z = 4t_0$ in both (a) and (b), $|\mathcal{\vec B}_{-}({\bf k}) |$ larger than 50 being plotted in red color; {(c)} Scaling relation between $S$ and $k_{\rm min}$, with parameters $t_z = 0.5t_{so}$ (upper, blue lines), $t_z = t_{so}$ (middle, red lines), and $t_z = 2t_{so}$ (lower, black lines). The triangle, solid, and dotted lines represent results in the continuous limit, lattice model, and asymptotic scaling lines.}
\end{figure}

Now we turn to the study of the scaling relation in a SO coupled WSM which can be realized by generalizing the 2D SO coupling proposed in recent work~\cite{LIU2} to 3D cubic lattice system. Note that this 2D SO coupling for Bose condensates has been realized in a recent experiment~\cite{LIUPAN}. We propose the 3D tight-binding Hamiltonian for the present WSM as
\begin{eqnarray}\label{eqn:tightbindingSI1}
H&=&-\sum_{<\bar{i},\vec{j}>}t_{\alpha}(\hat c_{\vec{i}\uparrow}^{\dag}\hat
c_{\vec{j}\uparrow}-\hat c_{\vec{i}\downarrow}^{\dag}\hat
c_{\vec{j}\downarrow})+\sum_{\vec{i}}m_z(\hat n_{\vec{i}\uparrow}-\hat n_{\vec{i}\downarrow})\nonumber\\
&&+\bigr[\sum_{j_x}t_{\rm so}(\hat c_{j_x\uparrow}^\dag\hat c_{j_x+1\downarrow}-\hat c_{j_x\uparrow}^\dag\hat c_{j_x-1\downarrow})+{\rm H.c.}\bigr]+\nonumber\\
&&+\bigr[\sum_{j_y}it_{\rm so}(\hat c_{j_z\uparrow}^\dag\hat c_{j_y+1\downarrow}-\hat c_{j_y\uparrow}^\dag\hat c_{j_y-1\downarrow})+{\rm H.c.}\bigr].
\end{eqnarray}
Here $t_\alpha$ denotes the spin-conserved hopping along $\alpha$ ($=x,y,z$) direction, $m_z$ and $t_{\rm so}$ represent an effective Zeeman term and spin-flip hopping coefficient, respectively~\cite{LIU2}.
Transforming $H_{\text{eff}}$ into momentum space yields $H = \sum_{k,\sigma, \sigma^\prime} \hat c_{\bf{k},\sigma}^\dagger \mathcal{H}_{\sigma,\sigma^\prime}(k) \hat c_{\bf{k},\sigma^\prime}$, with $\mathcal{H}(k) = 2 t_{so} \sin{(k_y)}\sigma_x +2t_{so} \sin{(k_x)} \sigma_y+(m_z - 2t_x \cos{k_x} - 2t_y\cos{k_y}-2t_z \cos k_z) \sigma_z$. For convenience, we consider the lattice to be isotropic in the $x$ and $y$ directions, so that $t_x=t_y=t_0$. Note that in the present system the TR symmetry, defined as ${\cal T}=i\sigma_y{\cal K}$, is broken.  In the parameter regime with $2|2t_0-t_z|\leq m_z\leq2(2t_0+t_z)$ and $m_z\geq2t_z$ (for $t_0,t_z>0$), we find that the above Hamiltonian has only two Weyl points located on the $z$ axis. In particular, for $m_z=4t_0$ the two Weyl points are given at $\bold k_{w1} = (0,0,\frac{\pi}{2})$ and $\bold k_{w2} = (0,0,-\frac{\pi}{2})$, with the Weyl cone Hamiltonian
\begin{eqnarray}\label{SOmodel}
\mathcal{H}_\pm(k) = 2t_{so}(k_y\sigma_x + k_x \sigma_y) + (t_0k_\perp^2\pm2t_zk_z) \sigma_z,
\end{eqnarray}
where $k_\perp^2=k_x^2+k_y^2$. The monopole Berry curvature nearby the Weyl points are shown in Fig.{\ref{figure2}}(b).

The band bottom of the system is located at $\bold k_0=(-\pi, -\pi,-\pi)$ [Fig.\ref{figure2}~(a)], where the BEC cloud can be accelerated toward one of the two Weyl points by a force in the $\hat e_{xy}-\hat e_z$ plane, with $F_x = (F/\sqrt{2}) \cos(\theta_0 + \theta)$, $F_y = (F/\sqrt{2}) \cos(\theta_0 + \theta)$, $F_z = F \sin(\theta_0+ \theta)$, where $\theta_0 = \arcsin(1/3)$. In the limit $\theta\rightarrow0$ the atomic cloud is accelerated to the Weyl point $(0,0,-\pi/2)$. The scaling relation in the regime $\theta\rightarrow0$ is shown to be $\ln S=-\ln k_{\rm min}+D_{\rm SO}$, with the intercept given by
\begin{equation}
D_{\rm SO}=\frac{1}{2}\ln\left| \frac{1}{9}+\frac{8t_{so}^2}{9t_z^2}   \right|.
\end{equation}
We present the numerical results in Fig.\ref{figure2}~(c), with different features being observed compared with those obtained in the previous LAT model. First, the intercept vanishes for $t_{\rm so}=t_z$ and it is independent of spin-conserved hopping $t_0$ in $x$ and $y$ directions. Secondly, due to the quadratic momentum term ($\propto k_\perp^2$), which is the leading correction to the Weyl cone Hamiltonian~\eqref{SOmodel} and absent in LAT model, the relation between the transverse drift $S$ and $k_{\rm min}$ approaches the linear log-log relation slowly. Nevertheless, we can see that this linear scaling relation can be well reflected when $k_{\rm min}<0.05$.

\begin{figure}[!t]
\centering
\includegraphics[width=0.95\columnwidth]{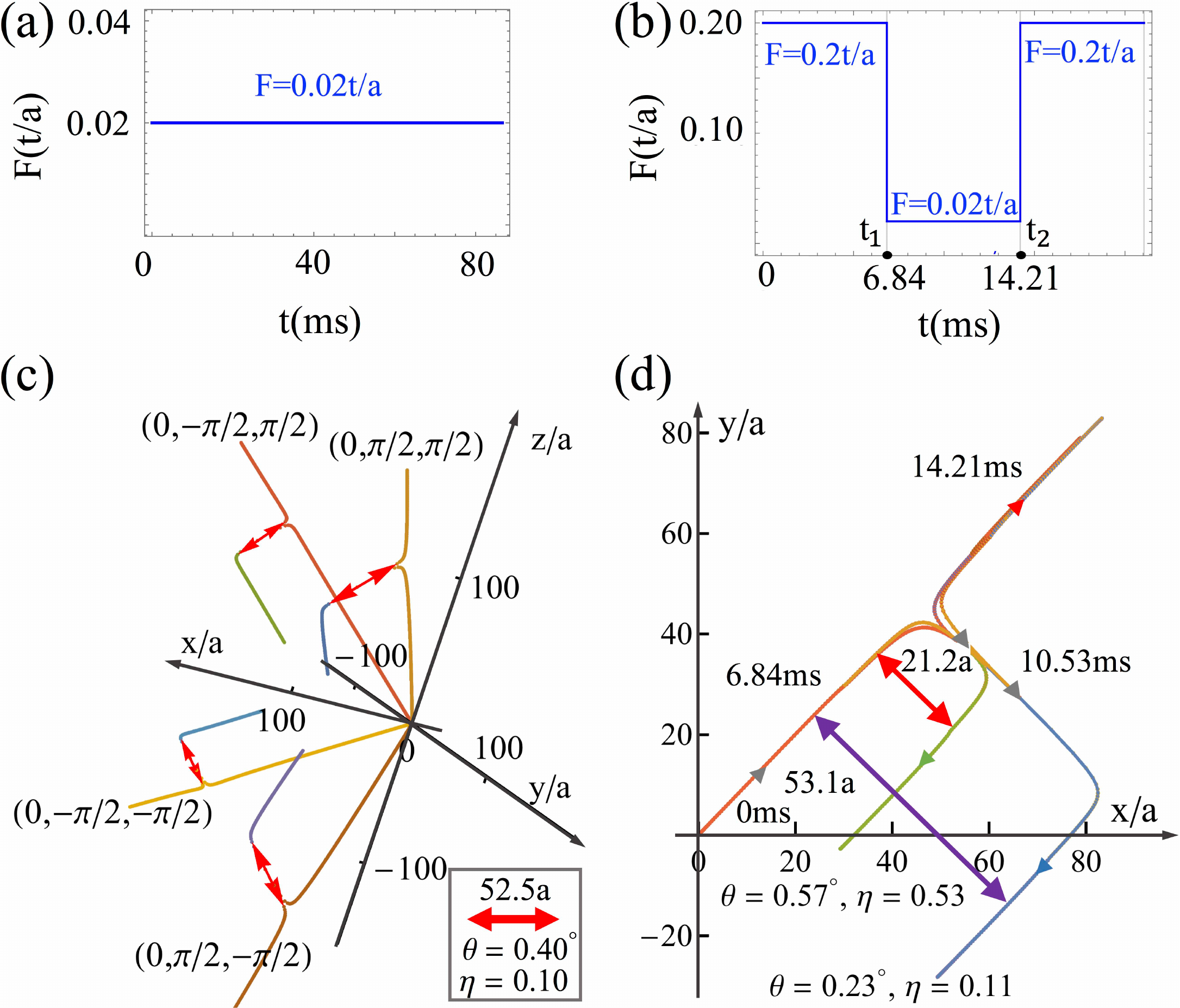}
\caption{ \label{figure3} {(a)} Manipulation of external force for WSMs in LAT model (a) and SO coupled model (b); {(c)} Trajectory of BEC cloud to four Weyl points in LAT Model, with $K_x=K_z=J_y = t = 2\pi \times$ 0.5 kHz. The LZ transition probability $P_{LZ} = 0.89$; {(d)} Trajectory of BEC cloud in SO coupled Model, with $t_{so} = t_z = t = 2\pi \times 0.52$kHz = 1.1$t_0$ and $m_0 = 4t_0$. The Lz transition probability $P_{LZ} = 0.89$ for $\theta=0.23^{\rm o}$.}
\end{figure}
{\it Landau-Zener transition.}--Note that when the BEC cloud is accelerated close to the Weyl point, the Landau-Zener (LZ) transition of the atoms can occur from the lower to higher subbands~\cite{LANDAU}. To ensure a sufficiently high resolution in the experiment, one requires that the ratio of the BEC cloud left in the lower subband after LZ transition should not be small. The LZ transition probability for a BEC cloud accelerated bypassing a Weyl point is $P_{\text{LZ}} = \exp(-\frac{\pi  E_{\rm min}^2}{F v_{\text f}})$, where $v_f$ is the magnitude of velocity when the BEC cloud approaches the Weyl point and $E_{\rm min}=2v_fk_{\rm min}$ is the energy difference between particle and hole states at the avoided crossing during the Bloch oscillation~\cite{LANDAU}. To measure the scaling behavior, we need to properly tune the force and its direction ($\theta$). First, a small angle $\theta$ and thus small minimal momentum $k_{\rm min}$ is necessary to ensure that the relation between the transverse drift $S$ and $k_{\rm min}$ can enter the linear scaling law regime. Secondly, a finite ratio of the BEC cloud is left in the lower subband after LZ transition so that the remaining atom cloud can be imagined. For fixed $k_{\rm min}$, though a small external force can supress LZ transition, it however leads to a large Bloch oscillation period which should not exceed BEC's life time. Thus proper manipulation of the force $F$ might be necessary.

For the WSM realized with LAT scheme, which has a relatively long life time, we consider a small constant force to observe the scaling behavior in the typical parameter regimes, as shown in Fig.~\ref{figure3} (a,c). The numerical simulation shows that more than $10\%$ of the BEC cloud remains in the lower subband after the atom cloud passes by the Weyl point, with the angle $\theta=0.4^{\rm o}$ or $k_{\rm min}\approx0.03$, which is well within the linear scaling regime according to Fig.~\ref{figure1} (c). On the other hand, for the SO coupled lattice model, to satisfy the aforementioned requirements we introduce a timing sequence for the manipulation of the external force. For the time $0<t<t_1=6.84$ms and $t>t_2=14.21$ms, a relatively strong force of magnitude $F_1=0.2t/a$ is applied, while within $t_1<t<t_2$ when the BEC cloud evolves very close to the Weyl point, a weak force of magnitude $F_2=0.02t/a$ is applied [Fig.~\ref{figure3} (b)].
With this manipulation we can show the ratio of BEC cloud left in the lower subband after LZ transition to be $\eta =1-P_{\rm LZ}=[1-\exp(-\frac{\pi  {E}(t_1)^2}{F_1 v_{f}(t_1)})][1-\exp(-\frac{\pi  {E_{\rm min}}^2}{F_2 v_{f}})]$. Here $E(t_1)$ and $v_f(t_1)$ stand for the energy difference between particle and hole states and the velocity at the time $t=t_1$, respectively. From the numerical results in Fig.~\ref{figure3} (d) we find that for an angle no less than $\theta=0.23^{\rm o}$ which corresponds to $k_{\min}=0.02$, the ratio of the remaining BEC cloud after LZ transition is over $10\%$ under the typical parameter regimes. Thus the linear scaling law can be observed with this configuration.

{\it Conclusion.}--We have predicted a linear logarithmical scaling law in Bloch oscillation dynamics in WSMs, and proposed that such scaling behavior can be applied to directly detect the Weyl nodal points in such topological semimetals. Being a consequence of the monopole structure nearby the Weyl points, the transverse drift of quasiparticles, which are accelerated bypassing a Weyl point, exhibits a linear log-log relation with respect to the minimal momentum measured from the Weyl point. This linear scaling behavior can provide a direct measurement of the complete information of Weyl points, including the chirality and anisotropy of the nodal points. Applying the present results to two lattice models for WSMs which can be experimentally realized in cold atoms, we have shown the feasibility of measuring topological Weyl points based on our prediction for future experiments.

This work is supported by NSFC and President's Fund for Undergraduate Research of Peking University.



\noindent

\begin{thebibliography}{99}

\bibitem{Wan2011}  X. Wan, A. M. Turner, A. Vishwanath, and S. Y. Savrasov,
Phys. Rev. B {\bf 83}, 205101 (2011).

\bibitem{Xu2011} G. Xu, H. Weng, Z. Wang, X. Dai, and Z. Fang,  Phys. Rev.
Lett. {\bf 107}, 186806 (2011).

\bibitem{Burkov2011} A. A. Burkov and L. Balents, Phys. Rev. Lett. {\bf 107}, 127205
(2011).

\bibitem{Hosur2012} P. Hosur, S. A. Parameswaran, and A. Vishwanath,  Phys.
Rev. Lett. 108, 046602 (2012).

\bibitem{Jiang2012} J.-H. Jiang, Phys. Rev. A {\bf 85}, 033640 (2012).

\bibitem{HASAN2} S.Y. Xu, I. Belopolski, N. Alidoust, {\em et al}. Science {\bf 349}, 6248 (2015).

\bibitem{DAIXI1} B.-Q. Lv {\em et al.}, Phys. Rev. X {\bf 5}, 031013 (2015).

\bibitem{DAIXI2} X. Huang {\em et al.}, Phys. Rev. X {\bf 5}, 031023 (2015).

\bibitem{HASAN} M. Z. Hasan and C. L. Kane, Rev. Mod. Phys. {\bf 82}, 3045 (2010).

\bibitem{QI} X.-L. Qi and S.-C. Zhang, Rev. Mod. Phys. {\bf 83}, 1057 (2011).

\bibitem{Nielsen1981} H. B. Nielsen and M. Ninomiya, Nucl. Phys. B. {\bf 185}, 20
(1981).
\bibitem{Nielsen1983} H. B. Nielsen and M. Ninomiya, Phys. Lett. B. {\bf 130}, 389
(1983).

\bibitem{Zyuzin2012} A. A. Zyuzin, A. A. Burkov, Phys. Rev. B {\bf 86}, 115133 (2012).

\bibitem{Son2013} D. T. Son, B. Z. Spivak, Phys. Rev. B {\bf 88}, 104412 (2013).

\bibitem{Parameswaran2014} S. A. Parameswaran, T. Grover, D. A. Abanin, D. A. Pesin,
and A. Vishwanath, Phys. Rev. X {\bf 4}, 031035 (2014).

\bibitem{Jiang2015} Q.-D. Jiang, H. Jiang, H. Liu, Q.-F. Sun, and X. C. Xie, Phys. Rev. Lett. {\bf 115}, 156602 (2015).

\bibitem{FZhang2015} S. A. Yang, H. Pan, and F. Zhang, Phys. Rev. Lett. {\bf 115}, 156603 (2015).

\bibitem{CZhang2015} C.-L. Zhang {\em et al}., arXiv preprint arXiv:1503.02630 (2015).

\bibitem{Xiong2015} J. Xiong, S. K. Kushwaha, T. Liang, J. W. Krizan,
W. Wang, R. Cava, and N. Ong, arXiv preprint arXiv:1503.08179 (2015).

\bibitem{Liu2009}X.-J. Liu, M. F. Borunda, X. Liu, and J. Sinova, Phys. Rev. Lett. {\bf 102}, 046402 (2009).

\bibitem{Lin} Y.-J. Lin, K. Jim{\'e}nez-Garc{\'i}a, I. B. Spielman, {\em Nature} {\bf 471}, 83-86 (2011).

\bibitem{Jinyi2012} J.-Y. Zhang {\em et al.}, Phys. Rev. Lett. {\bf 109}, 115301 (2012).

\bibitem{Wang} P. Wang {\em et al.}, Phys. Rev. Lett. {\bf 109}, 095301 (2012).

\bibitem{MIT} L. W. Cheuk {\em et al.}, Phys. Rev. Lett. {\bf109}, 095302 (2012).

\bibitem{Abeliangauge2} M. Aidelsburger {\em et al.}, Phys. Rev. Lett. {\bf 111}, 185301 (2013).

\bibitem{Abeliangauge3} H. Miyake, G. A. Siviloglou, C. J. Kennedy, W. C. Burton, W. Ketterle, Phys. Rev. Lett. {\bf 111}, 185302 (2013).

\bibitem{Abeliangauge4} M. Aidelsburger {\em et al.}, Nat. Phys. {\bf 11}, 162 (2015).

\bibitem{GJOS} N. Goldman, G. Juzeliunas, P. Ohberg, and I.B. Spielman, Rep. Prog. Phys. {\bf 77}, 126401 (2014).

\bibitem{HZhai2015} H. Zhai, Prog. Phys. {\bf 78}, 026001 (2015).

\bibitem{WULIU} X.-J Liu, X. Liu, C. Wu, and J. Sinova, Phys. Rev. A {\bf 81}, 033622 (2010).

\bibitem{Goldman2012} N. Goldman, J. Beugnon, and F. Gerbier, Phys. Rev. Lett. {\bf 108}, 255303 (2012).

\bibitem{Liu2013a} X. -J. Liu, Z. -X. Liu, and M. Cheng, Phys. Rev. Lett. {\bf 110}, 076401 (2013).

\bibitem{Vincent2013} X. Li, E. Zhao, and W. V. Liu, Nature Comm. {\bf 4}, 1523 (2013).

\bibitem{LIU2} X.-J. Liu, K.T. Law, T.K. Ng, Phys. Rev. Lett. {\bf 112}, 8, (2014).

\bibitem{Luming} S. -T. Wang, D. -L. Deng, L. -M. Duan, Phys. Rev. Lett. {\bf 113}, 033002 (2014).

\bibitem{Chuanwei2013} Y. Xu and C. Zhang, Phys. Rev. Lett. {\bf 114}, 110401 (2015).

\bibitem{Hickey2016} C. Hickey, L. Cincio, Z. Papi\'{c}, and A. Paramekanti, Phys. Rev. Lett. {\bf 116}, 137202 (2016).

\bibitem{HChen2015} H. Chen, X. -J. Liu, X. C. Xie, Phys. Rev. Lett. {\bf 116}, 046401 (2016).

\bibitem{KETTERLE} T.Dub\v cek, C.J. Kennedy, L. Lu, W.Ketterle, M. Soljacic, H.Buljan,  Phys. Rev. Lett. {\bf114}, 22  (2015).

\bibitem{HE} W. -Y. He, S. Zhang, K. T. Law	arXiv:1501.02348 (2015).

\bibitem{Alba2011} E. Alba, X. Fernandez-Gonzalvo, J. Mur-Petit, J. K.
Pachos, and J. J. Garcia-Ripoll, Phys. Rev. Lett. {\bf 107},
235301 (2011).

\bibitem{Price2012} H. M. Price and N. R. Cooper, Phys. Rev. A {\bf 85}, 033620
(2012).

\bibitem{LP} X. -J. Liu, K. T. Law, T. K. Ng, Patrick A. Lee, Phys. Rev. Lett. {\bf 111}, 120402 (2013).

\bibitem{Dongling2014} D. -L. Deng, S. -T. Wang, L. -M. Duan, Phys. Rev. A {\bf 90}, 041601(R) (2014).

\bibitem{LIU3} X. -J. Liu, Z. -X. Liu, K.T. Law, W.V. Liu, T.K. Ng, New J. Phys. {\bf 18}, 035004 (2016).

\bibitem{BD} A. A. Soluyanov, D. Gresch, Z. Wang, Q. Wu, M. Troyer, X. Dai, B.A. Bernevig, Nature, {\bf 527} 7579 (2015).

\bibitem{HCV} Y. Xu, F. Zhang, and C. Zhang, Phys. Rev. Lett. {\bf 115}, 265304 (2015).

\bibitem{EXTYPE21}J. Jiang et al., 
    arXiv:1604.00139.

\bibitem{LIUPAN} Z. Wu, L. Zhang, W. Sun, X. -T. Xu, B. -Z. Wang, S. -C. Ji, Y. Deng, S. Chen, X. -J. Liu, J. -W. Pan arXiv:1511.08170v1

\bibitem{LANDAU} C. Zener, Proc. R. Soc. London A {\bf 137}, 696 (1932); L.D.Landau, Phys.Z.{\bf 2},46 (1932).


\end{thebibliography}
\end{document}